# Inter-instance Data Impacts in Business Processes: A Model-based Analysis

Yotam Evron, Arava Tsoury, Anna Zamansky, Iris Reinhartz-Berger, Pnina Soffer

[1] University of Haifa, Mount Carmel, Haifa 3498838, Israel
{yevron,atsoury,annazam,iris,spnina}@is.haifa.ac.il

**Abstract**. A business process model represents the expected behavior of a set of process instances (cases). The process instances may be executed in parallel and may affect each other through data or resources. In particular, changes in values of data shared by process instances may affect a set of process instances and require some operations in response. Such potential effects do not explicitly appear in the process model. This paper addresses possible impacts that may be affected through shared data across process instances and suggests how to analyze them at design time (when the actual process instances do not yet exist). The suggested method uses both a process model and a (relational) data model in order to identify potential inter-instance data impact sets. These sets may guide process users in tracking the impacts of data changes and supporting their handling at runtime. They can also assist process designers in exploring possible constraints over data. The applicability of the method was evaluated using three different realistic processes. Using a process expert, we further assessed the usefulness of the method, revealing some useful insights for coping with unexpected data-related changes suggested by our approach.

***Keywords*** *– Business Process Model, Impact Analysis, Process Instances, Shared Data*

## 1. Introduction

In the past two decades, business process management (BPM) has become a leading discipline in industry and a focus of attention in academia. A main artifact in BPM is a process model, serving for representation, design, analysis, execution and mining of business processes. A process model depicts the expected behavior of a set of possible executions (instances) of the process. Many instances of the process may be executed, potentially in parallel, in runtime environments. These instances may share data and resources and interact through them. The interaction, however, is not represented in process models, and cannot be viewed, analyzed, or constrained through a model. In particular, when data value is changed in one process instance, the change may affect other instances and require propagating changes, potentially across additional process instances. Our aim in this paper is to support inter-instance impact analysis, focusing on impacts resulting from data shared across process instances. We note that while inter-instance impacts occur at runtime, analyzing shared data and resources is already possible at design time, based on process models, and implications of such analysis can be considered in advance.

Traditionally, process models have included information regarding control flow, neglecting data-related aspects. In recent years, data perspectives have been percolated and integrated into process models [9][10][12][16]. Particularly, process models that represent data operations have been suggested (e.g., PHILharmonicFlows [6] and Data Petri nets [9][13]). Furthermore, the impacts of data changes within processes have been investigated, leading to methods for data impact analysis [21]. These, however, are limited to impacts within a single process instance, leaving the analysis of inter-instance data impacts an open challenge.

To fill this gap, we build on the work presented in [21], which analyzed data impacts within a process instance, by suggesting how to analyze possible data impacts *across* process instances. Specifically, we study under which conditions additional process instances may be impacted by a change in a data value at one instance. We aim to generically address this question at design time, namely, to perform a model-based analysis when actual process instances do not yet exist [2]. This allows us to provide a general analysis independent of specific process instances and data values that may exist at runtime, thus to address all possible runtime impacts. The suggested method uses a data-aware process model (namely, a process model which includes data flows besides control flows) and a related data model, representing the data handled by the process. For simplicity of discussion and implementation, we assume that the data model is relational, i.e., composed of inter-related relations (tables). The outcomes of the method are potential inter-instance data impact sets, which may guide process users at run-time in tracking the impacts of data changes and supporting their handling. The obtained outcomes may also assist process designers in exploring possible constraints to be specified over data and designing exception handling mechanism to be triggered automatically upon runtime exceptions.

The remainder of the paper is organized as follows. Section 2 introduces a motivating example. Section 3 presents the formal foundations of our approach; Section 4 is devoted to the presentation of the underlying method. Section 5 presents and discusses the results of our evaluation studies. Finally, Section 6 discusses related work and Section 7 draws conclusions and presents future research directions.

## 2. Motivating Example

To illustrate the problem, consider a hotel booking process whose BPMN model is depicted in *Fig. 1*, presenting the activities of the process and its control flow. The process further uses data as activity inputs and routing constraints (for making path decisions at gateways), and creates data outputs. For example, the activity *room reservation* allocates a room for a specific customer; this activity is executed only if the type of room requested by the customer is available. This information is not explicitly shown in the figure.

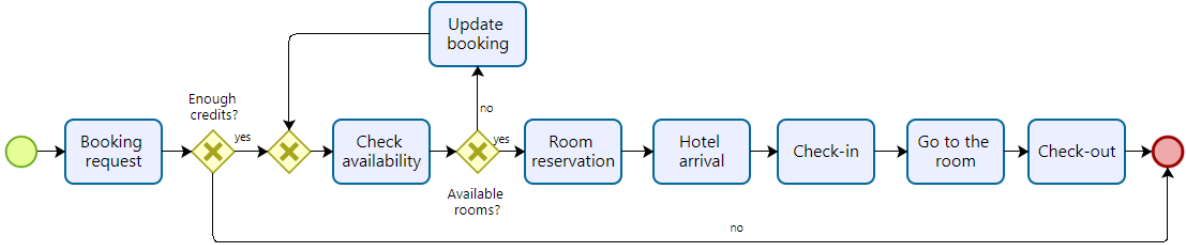

**Fig. 1**. Hotel booking process model

The data used and manipulated by processes is commonly stored in a database. A relational data model related to the hotel booking process is given online[1]. The model includes relations (tables) storing data about *Hotel-booking*, *Customer*, *Room*, *Hotel*, *Employee*, and more. *RoomRate*, e.g., is stored in the table *Room*, while *ReservationDate* and *TotalPrice* – in the table *Hotel-booking*. Additionally to the process and the data models, the organization has business rules, such as: (1) Booking a suite (a room

---

type) is only allowed if the *MembershipType* of the customer is gold/platinum. Moreover, gold/platinum members have a discount in their bookings according to their type of membership. The *MembershipType* is determined according to the number of rooms each customer has booked in the past year; (2) The *TotalCredit* set for a customer should cover the expected payment of bookings until payment is settled. A customer cannot place a booking that causes the *TotalCredit* limit to be exceeded.

(3) Each booking (*hotelBookingID*) can have only one room (*roomID*).

These business rules involve several process instances, which may be executed in parallel. Consider for example a situation where a customer who has several open bookings asks to add another room in one specific booking. Such a request may have a number of consequences. First, the *TotalPrice* would change, reducing the remaining credit of the customer. Second, the *MembershipType* can be upgraded according to the number of rooms the customer has booked in the past year. In case the *MembershipType* will change, as a result, the *TotalPrice* will require recalculation. Additionally, every active booking will have to be examined to determine whether the *TotalPrice* needs an update. This may affect not only the booking in which the change occurred but also other bookings which are reserved for the same customer.

The process model does not support identifying the full extent of such impacts and the required actions. To this end, we suggest an approach for analyzing possible data impacts across process instances, based on the available models, namely, the process and data models.

## 3. Formal Foundations

### 3.1 The data model

An underlying assumption for the proposed analysis is that all needed data is stored in a single relational database. We use standard database definitions [1], according to which a relational database DB is a set of relations $R \in DB$, where each *relation* R is a subset of the cartesian product over a set of attributes $Att_1, ..., Att_n$. A relation has *tuples* (rows), $r=(att_1, ..., att_n) \in R$, where $att_i$ is in the domain of $Att_i$, i.e., $att_i \in dom(Att_i)$. A tuple is uniquely identified through the primary key of the relation, $pk(R)$. Relations may be associated through foreign keys. We refer to associations among relations as mappings of their tuples – ref $\subseteq R_1 x R_2$. Table 1 lists common types of mappings, based on cardinalities among tuples.

**Table 1.** Types of cardinality between relations in a relational data model

| Cardinality | Notation | Definition |
|---|---|---|
| Many-to-many | m-m($R_1$,$R_2$)<br>m-m($R_2$,$R_1$) | Each tuple from $R_1$ may be mapped to more than one tuple in $R_2$ and vice versa, using mapping ref. Formally expressed:<br>• There are $r_1 \in R_1$, $r_2$, $r_2$'$\in R_2$ such that ($r_1$, $r_2$), ($r_1$, $r_2$')$\in$ref.<br>• There are $r_2 \in R_2$, $r_1$, $r_1$'$\in R_1$ such that ($r_1$, $r_2$), ($r_1$', $r_2$)$\in$ref. |
| One-to-many, many-to-one | 1-m($R_1$,$R_2$)<br>m-1($R_2$,$R_1$) | Each tuple of $R_1$ may be mapped to more than one tuple in $R_2$ and each tuple of $R_2$ can be mapped to at most one tuple of $R_1$, using mapping ref. Formally expressed:<br>• There are $r_1 \in R_1$, $r_2$, $r_2$'$\in R_2$ such that ($r_1$, $r_2$), ($r_1$, $r_2$')$\in$ref.<br>• For each $r_2 \in R_2$, there is at most one $r_1 \in R_1$ such that ($r_1$, $r_2$)$\in$ref. |
| One-to-one | 1-1($R_1$,$R_2$)<br>1-1($R_2$,$R_1$) | Each tuple of $R_1$ may be mapped to at most one tuple in $R_2$ and vice versa, using mapping ref. Formally expressed:<br>• For each $r_1 \in R_1$, there is at most one $r_2 \in R_2$ such that ($r_1$, $r_2$)$\in$ref.<br>• For each $r_2 \in R_2$, there is at most one $r_1 \in R_1$ such that ($r_1$, $r_2$)$\in$ref. |

### 3.2 Data-aware process models

In commonly used process modeling languages, such as BPMN, EPC, and Petri-nets, activities and control flows are the primary elements. Data and data flows may not explicitly appear or serve as secondary elements that are commonly under-specified. The following definition of a data-aware process model explicitly refers to data items and their integration in process models through input-output relations and routing constraints [21].

**Definition 1.** (data-aware process model): A *data-aware process model* is of the form M=(A, G, F, typeG, DI, RC, RF, IAO), where:

• (A, G, F, typeG) is a standard (control flow) process model; A, G, F are finite sets of activities, gateways, and control flows, respectively; F$\subseteq$(A$\cup$G)$\times$(A$\cup$G); typeG: G $\rightarrow$ {xor , and} is a mapping of gateways to their types.

• DI is a finite (non-empty) set of data items; a data item may be stored as an attribute in a relation, or temporarily used by the process.

- RC is a finite set of routing constraints, which are predicates over subsets of DI; we denote by support(rc) the set of data items related to the routing constraint rc (support (rc) ⊆ DI for each rc∈RC).

- RF ⊆ RC × F associates routing constraints in RC to control flows in F.

- IAO ⊆(DI ∪{null}) × A × (DI ∪{null}) is a set of data flows representing dependencies between data items and activities in the form of (input, activity, output)[2].

In our example (**Fig. 1**), *DI* includes all data items used by the process, such as *total amount*; *RC* includes the routing constraint [*TotalCredits > TotalPrice*], which is associated in *RF* with the flow between the XOR gateway labeled *enough credits?* and the activity *Check availability*; and IAO includes, among others, the tuples (*TotalPrice*, *check out*, *paid*) and (*null*, *booking request*, *check-in*).

We next define a process instance in this context.

**Definition 2** (Process instance): Let M be a data-aware process model. An *instance of M* is a (potentially partial) execution of the process model, which is represented as a triplet (caseID, σ, $\widehat{DI}$), where:

- caseID is the case identity, uniquely specifying an execution (i.e., for every pair of process instances, caseID$_i$≠caseID$_j$).

- σ=< $\widehat{a_1}, \widehat{a_2}$, …, $\widehat{a_n}$ > is a trace, i.e., a sequence of the activity instances performed in the execution, $\widehat{a_i}$ is an instance of $a_i$∈A. We say that $\widehat{a_i}$ *precedes* $\widehat{a_j}$, if i<j.

- $\widehat{DI}$ = { $\widehat{d_1}, \widehat{d_2}$,…, $\widehat{d_m}$ } is a set of data item instances that are used (as input or within a routing constraint) or modified (written as output) in the execution. Formally expressed, for every $\widehat{d_i}$ – an instance of $d_i$∈DI, the following holds: ($d_i$, $a_j$, $d'$)∈IAO, $\widehat{a_j}$∈σ is an instance of $a_j$, OR ($d'$, $a_j$, $d_i$)∈IAO, $\widehat{a_j}$∈σ is an instance of $a_j$, OR $d_i$∈support(rc), rc∈RC.

The notion of case identity bridges between data-aware process models and their related data models. Typically, a business process manages the lifecycle of a single object, which, conceptually, forms the identity of the process [22][24], and is often referred to as the *case object*. For example, for our hotel

---

booking process this object is *Hotel booking*; the case identities of hotel booking process instances are values of the primary key of the relation *Hotel booking*, namely, *hotelBookingID*. We thus call *Hotel booking* – the *identity relation* (IR) of the process and formalize it as follows.

**Definition 3** (Identity relation): Let M be a data-aware process model. The *identity relation* of M is a single relation $IR \subseteq Att_1 x \ldots x Att_n$, satisfying: (i) $\{Att_1,\ldots,Att_n\} \subseteq DI$, and (ii) for each instance PI of M, caseID of PI uniquely determines one tuple in IR and vice versa.

We note that process instances may read and write attribute values from additional relations besides IR, and these can be mapped to IR in different cardinality relations. If the cardinality of the mapping of some relation (R) to IR is many to one, m-1(R,IR), or one to one, 1-1(R,IR), then each tuple of R refers to exactly one process instance. In contrast, if the cardinality of the mapping is one to many, 1-m(R,IR), or many to many, m-m(R,IR), then every tuple of R may be "shared" by many process instances. This observation is crucial for our approach of identifying intra- and inter-instance impacts of data, which we describe next.

## 4. Intra and Inter-instance Impacts of Data

In this section, we present our approach which aims at identifying potential impacts across process instances, based on data-aware process models and their corresponding (relational) data models. The approach analyzes intra-instance data impacts (Section 4.1), defines inter-instance data impacts (Section 4.2), and suggests how to identify potential inter-instance data impacts (Section 4.3).

### 4.1 Intra-instance Data Impacts

As a process instance progresses, values of data items change and are used for determining the values of others. When the effects remain within a single process instance, we term them *intra-instance impacts*. These impacts can be analyzed based on the data flows (IAO) and the routing constraints (RC) specified in the process model, as suggested by [21]. A specific value of a data item might also be used along a chain of decisions and activities, affecting values assigned to other data items along the chain. The following two definitions formalize these notions.

**Definition 4.** (Model-level data impact): Let M be a data-aware process model, $d_1, d_2 \in DI$ and $act \in A$. $d_1$ has a *model-level* (direct) *impact* on $d_2$ with respect to *act*, denoted by $d_1 \rightarrow_{act} d_2$, if and only if one of the following holds:

- $d_1$ is used by *act* for writing $d_2$. Formally expressed, $(d_1, act, d_2) \in IAO$.

- $d_1$ is used in a routing constraint which leads to *act* that writes $d_2$. Formally expressed, $d_1 \in \text{support}(rc)$, and $\exists d'$ such that $(d', act, d_2) \in IAO$, where $rc$ is on a flow leading to *act*.

**Definition 5.** (Intra-instance data impact)*:* Let PI=(caseID, $\sigma$, $\overline{Dl}$) be a process instance of a data-aware process model M, $d_1, d_2 \in DI$. $d_1$ has *an intra-instance data impact on $d_2$ within process instance PI*, denoted by $d_1 \rightsquigarrow_{PI} d_2$ iff there is a sequence of model-level data impacts from $d_1$ that follows the order of activities in the process instance trace $\sigma$, and ends at $d_2$.[3] Formally expressed, $d_1 \rightarrow_{a0} d'_1$, $d'_1 \rightarrow_{a1} d'_2, \ldots,$ $d'_m \rightarrow_{am} d_2$, where $\{d'_i\}_{i=1..m} \subseteq DI$, $\{\widehat{a_i}\}_{i=0..m} \subseteq \sigma$, $\widehat{a_i}$ is an instance of $a_i$, $\widehat{a_j}$ precedes $\widehat{a_k}$ in trace $\sigma$ for all j, k such that j<k.

We denote by impacting(PI, $d_2$) the set of all data item instances d1' in PI for which $d_1' \rightsquigarrow_{PI} d_2$, and impacted(PI, d$_1$) – the set of all data item instances d2' in PI for which $d_1 \rightsquigarrow_{PI} d_2'$.

In our motivating example, *MembershipType* has a model-level data impact on *RoomType* (of *Room booking*) with respect to the activity *room reservation*. It also has an intra-instance data impact on *TotalCredits* through the activities *room reservation* and *check out* where the new amount of credits is updated, in process instances where these activities are performed.

**Listing 1** presents the pseudo-code for identifying potential intra-instance data impacts. Given a data-aware process model, the method returns an intra-instance impact set, intraSet, of all pair dependencies in the model that can lead to intra-instance data impact at runtime. The algorithm first identifies all the direct data impact pairs through activities (model-level data impact). Then, the algorithm seeks for impacts between a data item in the support set of a routing constraint and all the data items which are written in its specific path. Finally, the algorithm uses transitive closure for identifying the indirect impact between the data impact pairs that were found.

---

[3] Note that intra-instance data impact is reflexive, namely, $d \rightsquigarrow_{PI} d$.

---

**Input**: a data-aware process model **M**=(A,G,F,typeG,DI,RC,RF,IAO).

**Output**: **intraSet** – an intra-instance data impact set containing pairs (d, d') where d can have an intra-instance data impact on d'.

**Used methods**:

**reachableActivitySet(e)** – returns a set of activities which are reachable from model element e, which can be an activity or a routing constraint, e∈A∪RC

**transitiveClosure(set)** – returns the transitive closure of a set, e.g., by employing Warshall's algorithm [27]

**Begin**

intraSet ← { }

// model-level data impact through activity

for each (**d**,act,**d'**)∈IAO **do**

    intraSet ← intraSet ∪{(**d**,**d'**)}

// direct and indirect impacts through paths (routing constraints)

for each routing constraint **rc**∈RC  **do**

    for each data item **d**∈support(rc) **do**

        for each **act**∈**reachableActivitySet**(rc) **do**

            for each (**d''**,act, **d'**)∈IAO **do**

                intraSet ← intraSet ∪{(**d**, **d'**)}

// considering the transitive closure of intraSet

intraSet ← **transitiveClosure(intraSet)**

return intraSet

**End**      **Listing 1**. Pseudo code of the intra-instance data impact analysis

---

## 4.2 Inter-instance data impacts

At runtime, a change in the value of a data item may also have impacts on other process instances which are executed in parallel. This occurs through shared data items, which are stored in and retrieved from a single database [15] by different process instances. In our motivating example, the data item *roomID* of a specific room is shared by all the process instances whose bookings are associated with the same room. We first define *data sharing set at runtime* as sets of process instances sharing the same instance of a data item and may thus impact each other. We then generalize the definition to the indication of *data sharing sets* at design time, as functions over data item values.

**Definition 6.** (Data sharing set at runtime) Let M be a data-aware process model, DB – the associated database[4], IR∈DB – the identity relation associated with M. Let $d$ be a data item in DI in a relation $R_d$. For a process instance PI=(caseID, $\sigma$, $\widehat{DI}$), we denote by x the value of the primary key in the tuple $\hat{d}$ in $R_d$. The data sharing set at runtime of PI with respect to $d$ is defined by DS(d,x) = {(caseID, $\sigma$, $\widehat{DI}$) | $\widehat{PK(d)}$=x $\in \widehat{DI}$}.

For example, in our motivating example, the data item *roomID* of a specific room is a shared data item. The data sharing set DS(*roomID,15*) = {(caseID, $\sigma$, $\widehat{DI}$) | $PK(\widehat{roomID})$=15 $\in \widehat{DI}$}. In this specific example, room number 15 has a potential impact on all the process instances in this set where the value of the primary key in relation "Room" is 15.

**Definition 7.** (Data sharing set function) Let M be a data-aware process model, DB – the associated database, IR∈DB – the identity relation associated with M. Let $d$ be a data item in DI in a relation $R_d$. The data sharing set function: DSF(d) → all possible DS(d, x) for all possible values of x. At runtime, when $d$ and $x$ have specific values, DSF(d) returns the relevant DS(d,x).

For example, in our motivating example, the data sharing set function of the data item *roomID* is DSF(*roomID*) = {DS(roomID,1), DS(roomID,2), …}for each possible room id. At runtime, we can determine only the relevant data sharing set DS(d,x) as mentioned in Definition 6.

A data sharing set at runtime (DS(d,x)) may be trivial, namely, include a single process instance, or non-trivial (of a size greater than 1). According to Lemma 1, non-trivial data sharing sets may exist only for data items which are not part of the identity relation.

**Lemma 1**. |DS(d,x)| > 1 only if $d$ is not an attribute of IR (Identity Relation in Definition 3).

---

[4] We assume that {Att | Att is in relation R∈DB} ⊆ DI.

**Proof.** Assume $d$ is an attribute in IR. If DS(d,x) is non-trivial then there exist two different process instances PI, PI', such that $\hat{d} \in \widehat{DI} \cap \widehat{DI}'$, $\hat{d}$ is an instance of $d$, contradicting the uniqueness of case identifiers which refer to different IR tuples.

We rely on the concept of data sharing sets for defining *inter-instance data impacts*. As this concept refers to relations among process instances, it applies to runtime.

**Definition 8 .** (inter-instance data impact): Let M be a data-aware process model, DB – the associated database. Process instance PI=(caseID, σ, $\widehat{DI}$) has an inter-instance data impact on process instance PI'=(caseID', σ', $\widehat{DI}'$) through a data item d∈DI, denoted by PI ⇉d PI', iff:

- *PI, PI'* are in the same data sharing set of *d* (*PI, PI'* ∈ DS(d,x)).

- There is a data item $d_1$∈DI which has an intra-instance data impact on *d* with respect to *PI* ($d_1 \leadsto_{PI}$ d).

- There is a data item $d_2$∈DI such that *d* has an intra-instance data impact on $d_2$ with respect to *PI'* (d$\leadsto_{PI'}$ $d_2$).

- For *d*, there is $\hat{d}$ for which intra-instance data impact holds in both process instances. Formally expressed, $\hat{d}$ ∈ impacted(PI,$d_1$) ∩ impacting(PI',$d_2$).

In this case we say that *d* is a *shared data item*.

Note that it is possible that $d_1$ = d and $d_2$ = d.

For example, let us focus on the data item *MembershipType*. Its value is calculated according to the number of rooms the customer booked in the past year. Therefore, for each booking (*hotelBookingID*), there is an intra-instance data impact on *MembershipType* (*HotelBookingID* $\leadsto_{PI}$ *MembershipType*). In order to allow a customer to book a suite (*RoomType*), the *MembershipType* value should be gold/platinum. Thus, there is an intra-instance data impact on *RoomType* (*MembershipType*$\leadsto_{PI'}$ *RoomType*). We can conclude that data item *MembershipType* is a shared data item. Thus, the data sharing set at runtime of PI with respect to *MembershipType* is DS(*MembershipType*,x) = {(caseID, σ, $\widehat{DI}$) | $PK(\widehat{MembershipType})$=x ∈$\widehat{DI}$}.

### 4.3 Identification of Potential Inter-instance Data Impacts

Recall that our aim is to identify potential inter-instance data impacts at design time, taking into account all the possible process instances. Hence, we cannot enumerate the elements in data sharing sets, as defined in Definition 7. Yet, we can specify conditions under which process instances will be in a data sharing set for a data item $d$. We suggest using the associated data model and particularly the cardinality-based dependencies between the different relations and the identity relation of the process in order to identify the potential inter-instance data impact set, which is defined next. We then claim and prove that these sets provide a necessary condition for a data item to have inter-instance data impacts.

**Definition 9.** (potential inter-instance data impact set): Let M be a data-aware process model, DB – the associated database, $IR \in DB$ – the identity relation associated with M. The potential inter-instance data impact set is the set of all triplets (d1, d2; d) satisfying:

- $d_1, d_2, d \in DI$.

- $d$ is in relation $R \in DB$ whose cardinality with $IR$ is many-to-many or one-to-many. Formally expressed, m-m(R,IR) OR 1-m(R,IR).

- There is a possible process instance $PI$ of $M$ such that $d_1$ has an intra-instance data impact on $d$ with respect to PI. Formally expressed, $d_1 \leadsto_{PI} d$.

- There is a possible process instance $PI'$ of M such that $d$ has an intra-instance data impact on $d_2$ with respect to $PI'$. Formally expressed, $d \leadsto_{PI'} d_2$

The triplets that comprise a potential inter-instance data impact set can intuitively be described as follows. Given a process and a data model, there can be two instances of the process such that $d_1$ has an impact on $d$ in one process instance (PI); as $d$ is shared with the other process instance (PI'), any change in its value is propagated to PI', and can then have an (intra-instance) impact on $d_2$. This intuition is formalized in Theorem 1 and its corollary, that build on the concepts defined thus far for anticipating runtime impacts, based on properties that can be inferred at design time from the process and data models.

**Theorem 1**. Given a data item $d$ and two process instances PI, PI' at runtime. If PI≠PI' are in the same data sharing set of $d$, DS($d,x$), then $d$ is in a relation R∈DB whose cardinality with IR is m-m or 1-m.

**Proof.** Suppose for contradiction 1-1(R, IR) or m-1(R, IR). Then each tuple in R maps to at most one tuple in *IR*. Since a tuple in *IR* maps uniquely to a single process instance, caseID≠caseID' for any pair of process instances of *M*, particularly for *PI* and *PI'*. Hence DS($d,x$) is trivial and this is in contradiction to *PI ≠ PI'*.

The potential inter-instance data impact set provides the necessary conditions for inter-instance data impacts, as proven by the following corollary.

**Corollary.** The potential inter-instance data impact set contains all triplets ($d_1,d_2;d$) such that $d_1$, $d_2$, $d$∈DI and there are PI, PI' where $\widehat{d_1},\hat{d}∈\widehat{DI}$ in PI and $\widehat{d_2},\hat{d}∈\widehat{DI'}$ in PI', $\hat{d}$∈(impacted(PI,$d_1$)∩ impacting(PI',$d_2$)) in both PI and PI'.

**Proof**. Let ($d_1,d_2;d$) be a triplet such that there are PI, PI'∈ DS($d,x$) where $\widehat{d_1},\hat{d}∈\widehat{DI}$ in PI and $\widehat{d_2},\hat{d}∈\widehat{DI'}$ in PI', $\hat{d}$∈(impacted(PI,$d_1$)∩ impacting(PI',$d_2$)) in both PI and PI'. Since $\hat{d}$∈(impacted(PI,$d_1$)), $d_1\leadsto_{PI} d$ and since $\hat{d}$∈(impacting(PI',$d_2$)), $d\leadsto_{PI'} d_2$.

Since $\hat{d}$ is shared by both PI and PI' ($\hat{d}∈\widehat{DI}∩\widehat{DI'}$), both of them are in the same data sharing set of d (PI, PI'∈ DS($d,x$)).

By Theorem 1, $d$ is in a relation $R$∈DB whose cardinality with *IR* is many-to-many or one-to-many. Thus ($d_1,d_2;d$) is in the *potential inter-instance data impact set*.

Note that the potential inter-instance data impact sets provide necessary conditions and not sufficient ones, as actual inter-instance data impacts depend on runtime information not available in our model-based analysis, which only allows for design-time considerations. In our motivating example, the *total credit* of a customer, whose data sharing set is all the process instances that have the same *customerID*, may potentially have inter-instance data impact on data items such as *room-rate* for all instances in the data sharing set.

As the above discussion iterates between model-based concepts that relate to design time and concepts that apply to runtime, when process instances are created and operated, we summarize this distinction in **Table 2**.

**Table 2.** Summary of concepts for design time and runtime

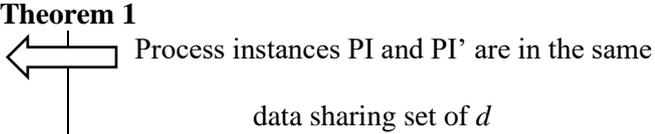

| Design time (model-based) concepts | Runtime (instance-related) concepts |
|---|---|
| Data-aware process model (Definition 1) Relational database model | Process instance (Definition 2) |
| Identity relation (Definition 10) ||
| Model-level data impact $d_1 \rightarrow_{act} d_2$ ( Definition 4): A direct impact of $d_1$ on $d_2$ via activity *act*. | Intra-instance data impact $d_1 \rightsquigarrow_{PI} d_2$ (Definition 5): An impact of $d_1$ on $d_2$ through a (partial) trace $\sigma$ in a process instance PI |
| Data sharing set function (Definition 7): A function which maps d to data sharing sets. Upon a given value of $d$ (available at runtime), a data sharing set will be returned | Data sharing set at runtime (Definition 6): the set of process instances that share a data item $d$ (have the same PK in the relation of $d$) |
| Potential inter-instance data impact set (Definition 9): A set of triplets ($d_1$, $d_2$, $d$) such that potentially PI $\rightrightarrows_d$ PI' may exist, where $d_1 \rightsquigarrow_{PI} d$ in PI and $d \rightsquigarrow_{PI'} d_2$ in PI' | Inter-instance data impact PI $\rightrightarrows_d$ PI' (Definition 8): PI has an impact on PI' through $d$ |
| **Theorem 1** |  |
| Cardinality between R of $d$ and IR is 1-m(R,IR) or m-m(R,IR) | Process instances PI and PI' are in the same data sharing set of $d$ |

For operationalizing the analysis, Listing 2 depicts the pseudo code for determining *Potential inter-instance Data Impact sets* (hereafter denoted as PDIs). It uses the intra-instance data impact analysis (Listing 1) and a DFS-like function, *findCardinality*, for retrieving the cardinalities between a given relation and the identity relation, and checks the condition for potential inter-instance data impact.

---

**Input**: a data-aware process model $M$=(A,G,F,typeG,DI,RC,RF,IAO), an associated database $DB$ and the identity relation $IR$ of the DB with respect to $M$.

**Output**: $PDI$ – the potential inter-instance data impact set containing triplets of the form $(d_1,d_2,d)$. Each triplet represents a potential inter-instance data impact for d.

**Used methods**: *findRel (d)* – returns the relation in DB in which $d$ is an attribute.

*findCardinality($R_1,R_2$)* – returns one of 1-1, 1-m, m-1, or m-m representing the cardinality between relations $R_1$ and $R_2$ in DB.

---

**Begin**

intraSet ← intraInstance(M)

PDI ← { }

// find an intra-instance data impact between $d_1$ to $d$

for each (**$d_1$, d**)∈intraSet **do**

    // find an intra-instance data impact between $d$ to $d_2$

    for each (d, $d_2$)∈intraSet do

        if (*findCardinality*(*findRel*(**d**), **IR**) is in {m-m, 1-m})

            PDI ← PDI ∪ {($d_1$, $d_2$, d)}

return PDI

**End**

**Listing 2.** Pseudo code of the potential inter-instance data impact set analysis (PDI)

After analyzing the potential inter-instance data impact sets, data item $d_1$ might have an impact on more than one data items in different triplets (various $d_2$). To support an analysis of all the possible impacts of any data item (as $d_1$) in the process, we introduce the following definition.

**Definition 10**. (impact set of $d_1$) Let M be a data-aware process model and PDI be the potential inter-instance data impact set which contains triplets of the form ($d_1$, $d_2$; d). Let $d_1$ be a data item and $D_1$ its data sharing set. The impact set of $d_1$ is the set of all the data sharing functions of d for all d such that $(d_1,d_2;d)$ in PDI for some $d_2$.

For example, let us consider the data item *membershipType*. This data item has a potential inter-instance data impact (discussed above after Definition 8) and thus has a potential inter-instance data impact set. In order to analyze the broader impacts, we need to analyze the impact of data items in PDI where they have impact on these potential inter-instance data impact sets too. In the PDI, *hotelBookingID* is a $d_1$ in one of the triplets while *membershipType* is the *d*. Therefore, the impact of $d_1$ (*hotelBookingID*) is a potential inter-instance data impact set which contains the potential inter-instance data impact set of *membershipType*.

---

**Input**: a data-aware process model $M$=(A,G,F,type,G,DI,RC,RF,IAO) and inter-instance impact set containing triplets of the form $(d_1,d_2,d)$.

**Output**: Aff-set($d_{aff}$) – the potential affected set of $d_{aff}$ containing all the data sharing functions which $d_{aff}$ might impact

---

**Begin**

For each data $d_{aff} \in DI$

    Aff-set($d_{aff}$) ← { }

    For each triplet $(d_1,d_2,d)$ where $d_1= d_{aff}$

        Aff-set ← Aff-set ∪ {data sharing set of d}

**End**

---

**Listing 3**. depicts the pseudo code for analyzing all the possible data-impact analysis for $d_1$ in the triplets.

In summary, the proposed approach enables a model-based analysis of inter-instance impacts that may materialize at runtime. Such indication can help process designers to anticipate possible runtime anomalies that may stem from these impacts and take appropriate measures.

## 5. Evaluation

We evaluated our approach in two different ways: First, we performed an applicability evaluation, where we applied the approach to three different process models (a production process[5] in a company in Israel, an extended order process[6] taken from the Northwind example [11] and a course opening management process[7] taken from an academic institution in Israel). The goal was to demonstrate our approach can handle the complexities of realistic processes, and the magnitude of the problems that are faced as a result of the impact of a change in a data item. Second, we performed a validation of the approach using a real-life production process with the help of a process expert. The goal was to validate the results obtained by our approach against a manual analysis performed by the process expert. The validation referred to runtime use of the model-based analysis, while the model-based analysis is a design-time analysis. As the goal was to validate the results, we could use our model-based analysis as a runtime solution to make sure the results are correct. Furthermore, we also asked the process expert to assess the usefulness of our approach and its possible contribution in practice.

For the evaluation phase we implemented the approach in a prototype system in an MS-SQL environment. Both data-aware process models and data models are stored in tables; the former – in tables separately storing the activities, the gateways, the routing constraints, the control flows and the data flows (IAO, see [21] for details).

### 5.1 Applicability Evaluation

We conducted an applicability evaluation using three realistic processes. Each one of them was analyzed using our MS-SQL environment. Table *3* provides details of the three processes. These include:

(1) a production process in a global company in Israel, which produces raw materials and special ingredients for various industries worldwide. The company has more than 250 employees. The production process, addressed in this study, is considered as a main process in the organization. The process includes the creation of a customer order, production of the items (shop floor production),

---

handling shipment to the customer, and receiving payment. Typically, hundreds of orders are handled in parallel.

(2) an extended order-to-cash process based on the Northwind Traders database [11]. We extended the basic database, adding relations (such as Shipping, Payment details etc.) in order to make our process model more comprehensive in terms of the main activities to be executed in a process instance.

(3) a course opening management process taken from an academic institution with more than 18,000 students. The process is operated using a SAP information system deployed in 2006. The process manages all the resources needed for opening a course in the academic system, including, lecturers, students, physical equipment needed during the classes etc.

Table 3. Quantitative summary of the processes' details

| Process | No. of activities | No. of routing constraints | No. of gateways | No. of relations in DB | No. of attributes (data items) | No of items in IAO |
|---|---|---|---|---|---|---|
| Production | 11 | 4 | 10 | 8 | 54 | 110 |
| Order-to-cash | 13 | 4 | 7 | 13 | 74 | 115 |
| Course management | 10 | 4 | 7 | 16 | 63 | 70 |

The goal of this study was to assess the applicability of the approach to different realistic processes and the magnitude of impacts that can be discovered by the proposed analysis. To this end, we analyzed the three processes using the prototype tool, examined the obtained output, and calculated the metrics given in Table 4 based on the PDI – the set of triplets $(d_1, d_2, d)$ that were identified for each process. The number of unique $d$ in PDI indicates the total number of data items that can be shared among instances in the process. The number of unique $d_1$ in PDI reflects the total number of data items that can trigger possible inter-instance impacts through a chain of intra-instance impacts that can affect a shared data item ($d$). In case there are inter-instance impacts in the process, we would like to know, in general, their consequences. As an indication, we calculate the average number of impact sets for a triggering data item $d_1$, and the average number of data items which might be affected ($d_2$). Last, we calculate the average number of $d_1$ for each $d$, which reflects the extent to which each shared data item ($d$) can be impacted within a process instance.

Table 4. Quantitative summary of the results concerning ($d_1$,$d_2$,d) triplets

| Process | Number of unique $d$ | Number of unique $d_1$ | Average number of impact sets for $d_1$ | Average number of $d_2$ per $d_1$ | Average number of $d_1$ per $d$ |
|---|---|---|---|---|---|
| Production | 5 | 10 | 2 | 2 | 2.8 |
| Order-to-cash | 4 | 11 | 5.91 | 1.45 | 4.5 |
| Course management | 3 | 4 | 1.5 | 1.25 | 1.67 |

According to Table 3, each process has a considerable number of data items (attributes) and input/output operations, which make the task of analyzing the impact of each data item a non-trivial task. Table 4 shows that the number of data items which might be involved in inter-instance data impacts does not depend on the number of data items handled in the process. For example, the average number of impact sets for $d_1$ in the course management process is 1.5, and in the production process it is 2 (4th column in the table), while the number of data items is 63 and 54 respectively. Therefore, there is no clear correlation between the number of impact sets and the number of data items in the process. Even data items that appear to have impacts only within a process instance may lead, through chains of intra-instance data impacts, to inter-instance ones. For example, in the order-to-cash process, each shared data item might be affected by 4.5 other data items (which are not shared with other instances) on average. This fact might render manual analysis of data items impact a complicated task.

In summary, the metrics presented above demonstrate how unpredictable analyzing an impact of a specific data item in a realistic process can be. Clearly, it is not a trivial task since there might be hundreds of data items or input/output operations. Furthermore, there is no correlation between the number of data items in the process and the number of impact sets in the process which will make it more difficult to predict the impact of a decision which was based on an inaccurate value.

## 5.2 Validity and usefulness evaluation

In this section, we evaluate our approach using a real-life case study. The case study is based on a production process[8] which was described in section 5.1 and is shown in Figure 2.

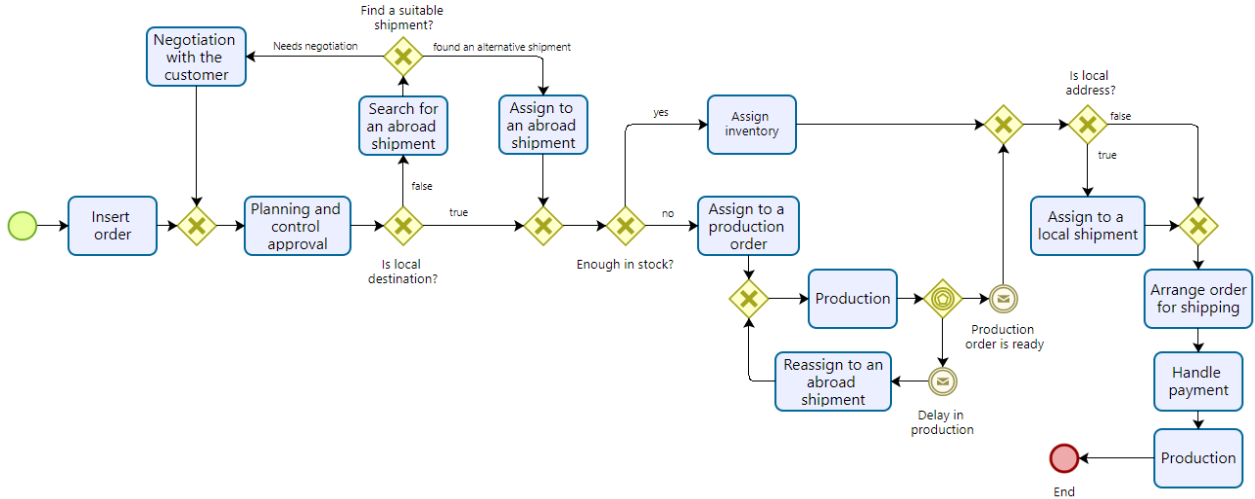

Figure 2. Production process

The production process is considered as a main process in the organization and includes the insertion of a new order, handing the shipment, managing the production of the items and receive the payment. We have obtained the company's permission to analyze this process with the assistance of the Production Planning and Control (PP&C) Manager. Using this case study as a validation of our approach, we followed the validation strategies proposed in [5] that incorporate domain experts.

Specifically, we employed the "engagement and understanding of the field" strategy when creating the process materials, and the "member checking" strategy for validating the results obtained by our approach. The first relates to the way the researchers can gain an understanding of the culture of the organization to interpret the data correctly. It requires the process expert selected for the study to meet the requirement of possessing the relevant knowledge and a central position in the organization. Our process expert is an Industrial Engineer who serves as a production, planning & control manager. The second strategy involves a confirmation of the validity of the results by the process expert, as described in detail next. We further conducted a reflection interview with the expert for assessing the usefulness of our approach.

---

[8] https://sites.google.com/view/shop-floor-process/home

### 5.2.1 Setting

The procedure of validating the results of our approach included the following steps:

1. <u>Conduct an introduction meeting with the process expert</u> – three semi-structured interviews of length 40-60 minutes were conducted with the process expert. In these interviews, we focused on the description of the process in general and of each activity, its input/output and the data items in the process. Moreover, we reviewed the database structure and relations. The interviews were recorded and analyzed.

2. <u>Model/Validate the process</u> – according to the information obtained from the interviews, we modelled the process described by the expert using BPMN, created the DB (database) and the respective IAO (a set of data flows representing dependencies between data items and activities). Then we presented the models to the process expert to validate them, making corrections as suggested by the expert where needed.

3. <u>Execute the algorithms</u> – our analysis algorithms were executed using the process and data models. The algorithms provided a list of potential inter-instance data impact sets which represent the sets of process instances that might impact each other. For example, in our motivating example, *total credit* of a customer is a data item whose data sharing set contains all the process instances that have the same value of *customerID* (the primary key of the relation).

4. <u>Conduct an in-depth summarizing interview with the process expert</u> – the interview had two parts: one devoted to validating our results (4.1) and the other to assess the usefulness of our approach (4.2).

    4.1 <u>Validation of the results</u> – We created five scenarios that include unexpected changes in data item values to be manually analyzed by the process expert (see Table *5*). These scenarios represent different kinds of possible inter-instance data impacts in the process. The expected results for each scenario were independently extracted from the results of our analysis (step 3). The impact of the described changes on various data items and process

instances was further discussed with the expert. The IR (identity relation) of the process is the *Orders*. For all the scenarios, we assumed there are many orders in the system.

Table 5. List of scenarios in the study

| No. | Scenario Description | Changed data item *d* | Relation of *d* | Cardinality with the IR (relation:*Orders*) | Number of process instances which might be affected | Expected inter-instance impacts (according to our analysis) |
|---|---|---|---|---|---|---|
| 1 | After an order (order no.12) has been assigned to a production order (production number no.55) and to a specific shipment (shippingID no. 122), it was realized during production (activity *Production*) that the estimated production date of the production order is not realistic and should be changed (change the estimated production date from 13.07.22 to 20.07.22) | Estimated production date | Production Order | 1:M | Many | All the process instances with the same production no. 55 and shippingID no. 122 will be impacted. |
| 2 | After assigning an order (order no. 7) to an abroad shipment (shippingID no.19 in activity *Assign to an abroad distribution line*), the customer asked to change the order's address to a local address. | Order address | Orders | 1:1 | Many (due to an intra-instance impact between orderAddress in Orders and areaDestination in Shipment.) | All the process instances with the same shippingID no. 19 will be impacted. |
| 3 | After assigning an order (order no. 15) to an abroad distribution (in activity *Assign to an abroad distribution line*), the customer decided to increase the quantity of a specific product (productID no. 8) to the order which has already been assigned | Quantity in order | Products In Order | M:1 | Many (due to (a) an intra-instance impact between QuantityInOrder in ProductsInOrder and UnitsInStock in Products, And (b) an intra instance impact between | All the process instances with shippingID no. 7889 and productID 8 will be impacted. |
| | to shipment (shippingID no.7889). | | | | QuntityInOrder in ProductsInOrder and ActualCapacity in Shipment.) | |
| 4 | After an order (order no.7) has been assigned to a shipment in activity *Assign to a local distribution line* (shipping no. 1782), transport prices increase due to an outbreak of the Covid epidemic. Thus, the shipping fee of the order (order no.7) changed. | Shipping fee | Orders | 1:1 | One | Only the process instance with order no.7 will be impacted. |
| 5 | While loading the orders (orders no. 9, 16 and 20) into the shipment (shipment no. 19), one of the employees found that the actual capacity of this shipment is not as recorded in the IS. | Actual Capacity | Shipment | 1:M | Many | All the process instances with the same shippingID no. 19. |

For each scenario, we started by presenting the process state. Then, we described an event resulting in a change in the value of a specific data item. The process expert was asked to indicate what might be affected as a result of the change that was made. Such indication was expected to rely on the expert's knowledge and experience, and simulate the way similar scenarios are handled as part of his daily work. Next, we showed him the results of our analysis and discussed together whether the results were compatible with his analysis. For impacts that were included in our analysis and not indicated by him, we asked whether they were correct, or falsely indicated. The analysis of each scenario took the process expert 15-20 minutes. This meeting was recorded as a remote interview.

4.2 <u>Assessment of usefulness</u> – After analyzing and discussing the scenarios as described, we asked the process expert to express his opinion about (1) the information that our approach is able to provide, and specifically – to what extent this information would be useful, (2) how he would use such information if it becomes available to him, and (3) what kinds of problems could be solved by using the approach.

### 5.2.2 Results

As mentioned above, the process expert discussed the potential impacts of each scenario and the matching results are shown in **Table 6** . For each scenario, we assumed there are many orders running in parallel.

**Table 6.** Results of each scenario validation

| Scenario no. | Changed data item | Manually identified by the domain expert (yes / partially) | Validated by the domain expert (yes/no) | Comments |
|---|---|---|---|---|
| 1 | Estimated production date | Partially | Yes | The process expert did not indicate the impact on the shipping date. After a short explanation, he agreed the impact was correctly spotted |
| 2 | Order address | Yes | Yes | |
| 3 | Units in stock | Yes | Yes | |
| 4 | Shipping fee | Yes | Yes | |
| 5 | ActualCapacity | Yes | Yes | |

Our analysis results fully matched the process expert's specification, except for the first scenario, where the expert did not specify the full impact that had been identified by our approach. More specifically, according to our analysis, if the expected production date is delayed, the shipping date might also be impacted, and this was omitted by the expert. However, when discussing the output of our analysis, he agreed that this impact was also possible and that it was overlooked by him.

In conclusion, the process expert validated all our model-based predictions. Moreover, the one difference between our results and the process expert's results was where he failed to identify the full impact (which he later approved to exist). This shows that manual analysis, even by a process expert who identifies data impacts in the specific process daily, is error-prone and an automated analysis is needed.

## 5.3 Reflection on the approach

After the discussions regarding the scenarios, we asked the process expert to reflect on the results provided by our approach, addressing specifically the following issues.

### 5.3.1 Information the approach can provide

The expert's opinion was that the impact information is extremely useful for situations where a decision needs to be made following an unexpected change in the value of a data item. The expert said "in organizations where the relations in the DB are complicated, for example – a specific shipment can involve several orders, the approach can be useful to support analyzing the impacts among process instances". Moreover, the number of possible actions that should be considered in response to an unexpected change and their indirect impacts on other process instances might be difficult to analyze, depending on the complexity of the DB.

### 5.3.2 Potential uses such a system can have

The process expert said that currently one cannot be certain the manual analysis is correct. For that reason, our approach can help verify the expert analysis or support it. According to him, usually when a possible data change occurs, the process expert needs to calculate and plan possible future actions that are needed due to the change. This predictive analysis phase is a difficult task and it is done manually in the organization. Therefore, to be able to mitigate the risk of errors in this challenging task, the current policy is of producing 20% more than required for each order to avoid relying on an analysis of future impacts, which is unreliable with their existing tools. If impact analysis is automatically and reliably available, this policy can be changed, and the result would be a significant reduction of costs.

### 5.3.3 Types of problems such a system can assist in solving

According to the process expert, besides the support discussed above, additional benefits can be gained. In particular, impact information can assist the marketing department in negotiating with the customers about delays or any changes in their orders. Currently, marketing people make decisions during this negotiation without any further information about possible impacts on additional process instances. Having this information available might reduce negotiation time and improve the results, as impacts can be considered even before consulting the production planning manager. Moreover, the availability of such information might lead to additional courses of action. For example, it might help the marketing department decide they can reallocate stock from one order to another following urgency considerations, with a full view of the impacts of such action.

## 5.4 Discussion of the results

In the first phase of the evaluation, we showed that our method can analyze processes with various degrees of complexity. These realistic processes might include many data items and some complex relations between data items among process instances. For example, we demonstrated that there is no correlation between the number of data items and the number of impact sets in the process. This shows that it is a difficult task to, e.g., predict a data impact of a specific decision which was based on an inaccurate value, particularly if it is done manually. Furthermore, the analysis of inter-instance data impacts requires tracking the impacted data items within the instance (intra-instance), and not only data items that are shared among process instances. Thus, we can conclude that analyzing the data impacts in real life processes is not an easy task to perform manually, and support is needed for this task. Our proposed method provides such support by identifying these impacts automatically. Generalizability of the approach has been demonstrated by gaining the results for different types of processes.

In the second phase of the evaluation, we validated the results of our method in a case study and obtained an expert assessment of its usefulness. In four of the five scenarios we examined the impacts identified using our algorithms were also identified manually by the domain expert. In the fifth scenario our approach identified impacts that were not spotted manually by the expert, but were validated when specifically introduced to him. These findings provide a full validation of our results in the case study, and  demonstrate the error-prone nature of this task when performed manually.

When we asked about the usefulness of our approach, it was clear from his answers that such a tool can be very helpful for process designers. For example, analyzing the impact of an incorrect data item which is used in a decision (e.g., quantity in stock). Moreover, it can assist in exploring what might happen in case of a change in one data item at runtime, and to identify the full impacts of such a change on other data items and process instances. He also stated that the availability of such a tool can eventually contribute to reduction of costs which are currently incurred by extra-production, intended to mitigate risks of dynamic changes and the associated complicated decisions. In the studied company, he agreed, this may be possible because the same product can be sold in different orders. Nevertheless, in general, his view is that tool support for handling dynamic changes is essential.

We note that the case study referred to a runtime use of the proposed analysis, although the analysis is model-based an can primarily be used at design-time. While the case study can be considered as showing usefulness of the analysis in this context, we argue that it will further be useful when used at design time. Having such an analysis while designing the model can assist in predicting and understanding the process better, and may lead to changes in the process design to mitigate risks of dynamic changes at runtime (e.g., adding control steps of shared data items). Such early design decisions might reduce expensive consequences of dynamic changes at runtime. Moreover, as the analysis is model-based and refers to stable structures of the process, its results can support runtime, as shown in the case study, and have additional uses (as indicated – for negotiation of possible changes with the customer).

## 6. Related work

Business process models typically describe the behavior of a single instance, isolated from others. Recently more works address relations and interactions among process instances. We review these works in three different phases: design time, runtime and process mining based on event logs.

Design time analysis of interactions among process instances relates to specifying and modeling such interactions and analyzing them based on these models. Runtime management and execution of such interaction is mostly based on design time models that encompass this behavior.

In [7], an approach for specifying instance-spanning constraints was suggested. This is further developed by Fdhila et al. [3], where, based on an extensive collection of reported constraints, formally specified generic constraint patterns were developed. These works do not consider or specify the set of instances to which a specific constraint/relation apply.

Another example of analyzing interaction between processes is the approach reported in [12], where the interaction is formally analyzed in the context of a possible verification of the process model. The authors show that models that allow reading and writing shared data are difficult to analyze, and that decidability of these constraints is only possible when certain limitations are imposed over available data operations. This formal analysis supports and motivates our work, demonstrating the complexity of managing shared data situations. In [20], the authors try to tackle the notion of process relation awareness at design time and runtime. They suggested the relational process structure that allows to track many-to-many relationships and enforce cardinality constraints.

The Object-Centric Behavioral Constraint (OCBC) model [8][23] is an example modeling approach which combines a process model with a data model into one new modeling language, where the modeler is able to add behavioral constraints related to data using the run-time notion of event logs in order to check conformance with the specified constraints.

All of these works define new models or specifications in order to tackle the notion of interaction among process instances. Our approach, in contrast, relies on a combination of existing (traditional) models of process and data. We extract information from each of the models so that we can better understand the relationship among the process instances. The fact that we use existing models makes the approach generic and easier to implement in cases where these models already exist.

While our approach is model-based, its in-depth analysis of interactions among process instances can also contribute to the development of process mining approaches where such interactions are of relevance, by informing them on relevant interactions among process instances.

In particular, relevance of interaction among process instances (cases) has been recognized as relevant in the context of prediction tasks. Examples include Senderovich et al. [18], who presented an adaptive method for predicting the remaining time of a process instance on the basis of both intra-case and inter-case properties that can be found in a process event log. In [4], the authors suggested a conceptual framework for predicting the completion time of a process instance, considering inter-case properties that relate to control-flow, time, resources and data. These papers show that an explicit use of the inter-case properties can significantly improve the prediction results.

Additional process mining analyses that apply to interaction among process instances include queue mining [17] and its consideration for time prediction [19], as well as discovery and analysis of batching behavior, [26][28]. While in queuing and batching the interaction among process instances is centered around resources, it may also be reflected in data-related interactions.

These process mining approaches show the importance of understanding the relationship among process instances to be able to improve the processes. As process mining solutions, they use runtime knowledge such as event logs, while in our work we aim to identify such interactions between process instances already during design time.

One additional relevant area related to process mining is the recent trend of analyzing object-centric event logs, as opposed to the case-centric ones used so far. In [25], a new event log notion is proposed where events can refer to any number of objects. [14] proposed an approach using the object-centric event log (OCEL) to calculate performance measures in the object centric setting such as concurrency, loops etc. Moreover, their approach supports the interaction between objects, e.g.,, considering the synchronization time among them. While this direction is in-line with our approach of broadening the view beyond a single process instance, it may become unclear what should be considered a process instance. The analyses supported so far by object-centric approaches relate to behavior along the life-cycle of a single object (as directly-followed graphs) or to interaction among objects. Interaction among different instances of the same object have not, to the best of our knowledge, been addressed so far. We hence suggest that object-centric process mining can gain from our inter-instance data impact analysis, since it opens a direction which is yet to be addressed in this context.

In summary, the relevance of going beyond the view of a single process instance has been recognized, but not fully accomplished so far. Our work's contribution in this context is to give a generic view and a possibility for design time analysis that can identify the possible inter-instance data impacts sets. Moreover, combining intra-instance and inter-instance data impacts has not, to the best of our knowledge, been suggested so far.

## 7. Conclusion and Future Work

The paper suggests a design-time approach for identifying inter-instance data impacts based on data-aware process models and the corresponding data models, without runtime information. This is done by finding potential inter-instance data impact sets in the form of triplets that depict the dependencies between affecting data items and shared data items. By this, we contribute to the understanding of interactions that take place among process instances. An additional contribution is the combination of intra-instance and inter-instance data impacts, which allows assessing both direct and indirect consequences of changes in data values, within a process instance and across process instances.

While the paper uses a composition of a data-aware process model and a (relational) data model, the notions introduced are generic and independent of a specific data or process modeling formalism. The main conceptual idea, which is based on activity inputs and outputs and cardinality-based relations, combined with intra-instance data impacts, is notation-independent.

Although our approach is model-based and uses design time models, it can be applied both at design time and at runtime. At design time, the approach can support the specification of structural constraints among process instances in two ways. First, it can help assess the scope of applicability of each data sharing set, which would then become explicit in the specification of constraints. Second, it can illuminate possible impacts that may need to be constrained, thus evoke the creation of additional dependencies. At runtime, the approach can assist in identifying the affected process instances in case of exceptions, ad-hoc changes, deviations, and unexpected changes in data values. These situations require careful actions to be taken by users for handling the consequences. Our approach can be considered as a first step towards providing useful guidance and support for these tasks. Furthermore, as suggested in our case study evaluation by the process expert, awareness of the possible impacts of a change can play an important role in deciding whether to make such a change. Specifically, the process expert suggested that marketing people, who interact with the customers and negotiate with them may use the proposed analysis approach for making informed decisions considering their full impacts.

Our approach focuses on design time, but can be applied during runtime too. One limitation of our approach for runtime use is that it does not consider the state at which process instances are at the moment of data item's value change (runtime). While our method finds data sharing sets, it is not necessary that all process instances in those sets would indeed be affected, as some may be at states where the impact are irrelevant (e.g., orders that are already on the truck). Extending the approach to cater for this aspect is one of the future research directions. Another direction is to consider impacts within paths of execution. This will allow for further analysis of impact chains relevant to given process instances, considering their effects to be confined within the relevant process paths. Further evaluation of the approach is also planned, using additional processes to assess generality and validating the results in real life settings.